# Investigation of inter-grain critical current density in $Bi_2Sr_2CaCu_2O_{8+\delta}$ superconducting wires and its relationship with the heat treatment protocol

I. Pallecchi [1,†], A. Leveratto [1,†], V. Braccini [1], V. Zunino [1,2] and A. Malagoli [1,*]

[1] National Research Council CNR-SPIN Genoa, C.so Perrone 24, 16152 Italy

[2] Department of Physics, University of Genoa, via Dodecaneso 33, 16146 Italy

[†] These authors equally contributed to the work

* andrea.malagoli@spin.cnr.it

**Abstract**

In this work we investigate the effect of each different heat treatment stage in the fabrication of $Bi_2Sr_2CaCu_2O_{8+\delta}$ superconducting wires on intra-grain and inter-grain superconducting properties. We measure magnetic critical temperature $T_c$ values and transport critical current density $J_c$ at temperatures from 4 K to 40 K and in fields up to 7 T. From an analysis of the temperature dependence of the self-field critical current density $J_c(T)$ that takes into account weak link behavior and proximity effect, we study the grain boundaries (GB) transparency to supercurrents and we establish a relationship between GB oxygenation in the different steps of the fabrication process and the GB transparency to supercurrents. We find that grain boundary oxygenation starts in the first crystallization stage, but it becomes complete in the plateau at 836 °C and in slow cooling stages, and is further enhanced in the prolonged post annealing step. Such oxygenation makes GBs more conducting, thus improving the inter-grain $J_c$ value and temperature dependence. On the other hand, from the inspection of the $T_c$ values in the framework of the phase diagram dome, we find that grains are oxygenated already in the crystallization step up to the optimal doping, while successive slow cooling and post annealing treatments further enhance the degree of overdoping, especially if carried out in oxygen atmosphere rather than in air.

**Introduction**

Superconductors with a high critical temperature such as high-$T_c$ cuprates have potential application in the superconducting power cables and high field magnets. $Bi_2Sr_2CaCu_2O_{8+\delta}$ (Bi-2212) superconducting round wires have been indicated as the best conductors for magnet technology for fields well above 30 T [1,2]. Indeed, despite having no macroscopic texture and containing high angle grain boundaries, they sustain very high superconducting critical current densities $J_c$ of 2500 A/mm$^2$ at 20 T and 4.2 K [1]. Most remarkably, the round cross section is particularly appealing in that it is anisotropy-free and it facilitates magnet design. Square or rectangular shaped cross sections in a solenoid type winding allow a better compactness of the turns and reduce the void space between them. Smooth corners of such cross section avoid damage in the insulation. Moreover, with a squared wire it is possible to avoid unwanted conductor twisting during the winding process. However, as compared to round wires, uniformity of the cross section is more critical for squared wires. Recently, it was demonstrated that the round shape induces a local biaxial texture (FWHM <15°), which in turns determines an azimuthal rotation of the grain c-axes along and about the filament axis, thus generating macroscopically isotropic behavior and large values of $J_c$ [3]. Yet, to further maximize inter-grain current in Bi-2212 wires, the main issues to investigate and improve are the presence of voids due to bubbles grown during the melt stage and the formation of secondary and off-stoichiometric phases within filaments (mainly either $Bi_2Sr_2CuO_x$ (Bi2201) grains or Bi-2201 intergrowths within Bi-2212 grains) and at the grain boundaries. The former problem, which is responsible for the degradation of the supercurrent over long lengths, can be addressed either by applying an over pressure during the heat treatment [1,4,5] or by the use of a suitable alternation of groove rolling and drawing to diminish the porosity prior to the heat treatment [6]. On the other hand, the latter problem requires fine tuning of fabrication protocol, and there may still exist an edge of improvement in this direction. On the individual filament scale, secondary phases such as Bi-2201 grains obstruct the supercurrent transport, by reducing the Bi-2212 grain connectivity [7,8]. The effect of oxygen post annealing on secondary phase formation and microstructure has been studied in [9]. In YBa$_2$Cu$_3$O$_{7-\delta}$ (YBCO) conductors, transparency of grain boundaries was improved by chemical (Calcium) doping [10], which heals their small carrier density and proximity to a parent, antiferromagnetic insulating state at the grain boundaries. Further investigations in this sense clarified the role of non-uniform segregation of Ca ions on a scale of ~1 nm near the dislocation cores, driven by the local strain and charge imbalance at the grain boundaries [11]. However a more recent systematic study of grain boundary transparency in YBCO films on bicrystal substrates as a function of tilted magnetic field, temperature, Ca-doping level and oxygen content evidenced that the effectiveness of Ca doping is limited to temperatures close to T$_c$ and for fields perpendicular to the grain boundary plane [12]. In Bi-2212 an analogous beneficial effect of doping was obtained by means of low temperature oxygenation treatments, that overdope the Bi-2212 phase in ways that are generally not possible in (Bi,Pb)$_2$Sr$_2$Ca$_2$Cu$_3$O$_x$ (Bi-2223) and YBCO [13].

Systematic and extensive characterization of fabricated Bi-2212 wires is an essential part of the optimization of the fabrication protocol. Moreover, the measurement of the critical current in a wide range of magnetic fields and temperatures does not only provide a feedback for the fabrication process, but it also sets the limits of application in the relevant field and temperature regimes and allows investigation of intrinsic and extrinsic superconducting properties through data analysis and comparison with modelling. In particular, the temperature dependence of J$_c$ has been poorly addressed so far, indeed the few works reporting J$_c$(T) data [14,15,16] mostly date back to decades ago, and are obtained on wires of modest performance prepared with different procedures as compared to those obtained at present [14,15].

In this work, we present a full characterization of transport J$_c$ behavior as a function of temperature from 4 K to 40 K and magnetic field up to 7 T in Bi-2212 wires prepared with different protocols, purposely chosen to investigate the effect of oxygen overdoping on the grain boundary transparency to supercurrent and on the superconducting properties in general. As an independent estimation of the inter-grain current, we also carry out measurements of remnant magnetization. We analyze the transport J$_c$(T) dependence, extracting information about the weak link behavior and proximity effect at the grain boundaries. We also inspect J$_c$(T,B) curves and apply collective pinning scaling laws to probe the flux pinning dynamics. We find that grain boundary oxygenation starts in the first crystallization stage, but it becomes complete in the plateau at 836 °C and in slow cooling stages. A further grain boundary oxygenation occurs in the prolonged post annealing step.

**Experimental**

**a. Characterization techniques**

The microstructure and elemental composition of our Bi-2212 wires is inspected by detection of secondary emission using a Scanning Electron Microscopy (SEM) and by Energy Dispersive X-ray (EDX) analysis. Further characterization is carried out by magnetization experiments in a commercial

Magnetic Properties Measurement System (MPMS, by Quantum Design) with magnetic field parallel to the wire axis, and by 4-probe transport critical current experiments in a self-made apparatus described in detail in [17], at temperatures from 4 K to 40 K and in magnetic fields perpendicular to the wire axis up to 7 T. As for magnetic measurements, we carry out zero field cooled dc susceptibility measurements in a 10 Oe field to extract the critical temperature $T_c$ at the onset of the diamagnetic superconducting signal and remnant magnetization measurements to separate inter- and intra-grain contributions to the magnetic critical current density. In the latter case [18,19], the sample is subject to incrementally increasing magnetic field $H_{max}$ parallel to the wire axis, and remnant magnetization $m_R$ is measured for each $H_{max}$ run, after removal of the applied field. Differentiation of the measured $m_R(H_{max})$ curves yields two peaks, and each peak can be associated to a product of screening current and length scale of this current. The lower field peak is associated to magnetic flux first entering grain boundaries weak links, and thus the value of $m_R$ at this lower field peak allows to extract the inter-grain magnetic critical current density, once the macroscopic sample geometry is taken into account.

**b. Sample preparation**

In this work we present characterization of a series of Bi-2212 wires prepared with different protocols, aimed at exploring the role of intra-grain and grain boundary oxygen doping in determining the superconducting properties and the inter-grain critical current density.

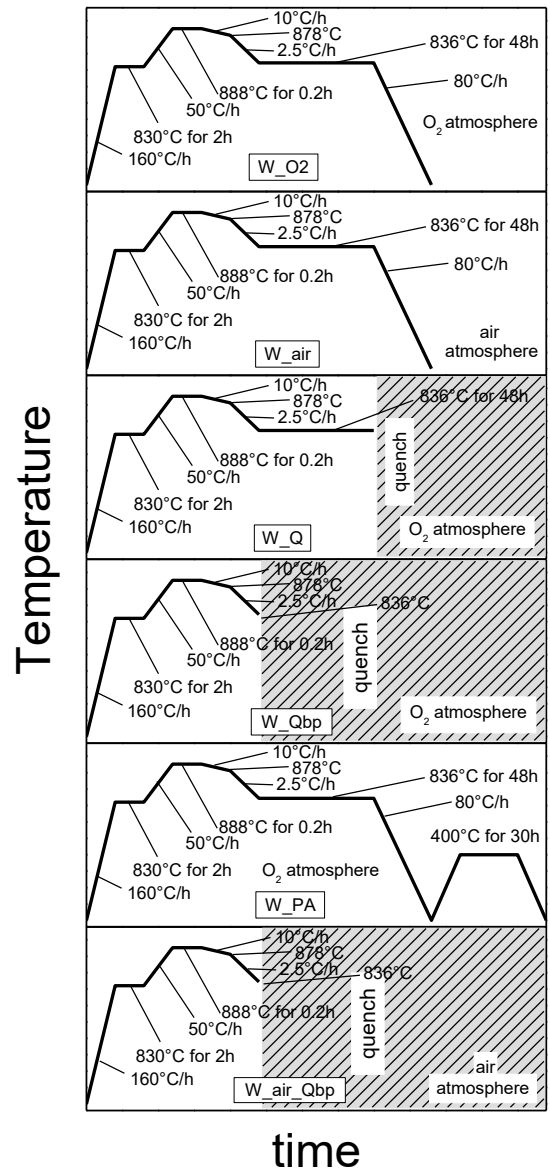

**Figure 1: Temperature treatment protocols**. Sketches of the temperature treatment protocols of the samples presented in this work. Both horizontal and vertical axes are in arbitrary units and not in scale, for clarity sake.

All the samples are prepared using the powder-in-tube method, starting by filling an Ag tube with outer (OD) and inner (ID) diameters of 15 and 11 mm, respectively, with Nexans granulate powder of composition $Bi_{2.16}Sr_{1.93}Ca_{0.89}Cu_{2.02}O_x$, as described in [6]. After drawing, the obtained hexagonally-shaped monofilamentary wire is cut in 85 pieces, restacked in a second 12.5/11 (OD/ID) mm Ag tube and then cold worked with a proper alternation of drawing and groove-rolling steps. The obtained wire, which is again hexagonally-shaped, is cut into 7 pieces and restacked in a 9.5/8 mm (OD/ID) Ag/Mg alloy tube. The restacked tube undergoes again a process with alternation of drawing and groove-rolling. Finally a square wire consisting of 7 bundles of 85 filaments each is obtained, having a 0.7 x 0.7 mm$^2$ cross-section shaped as a square with smooth corners and a superconducting fill factor of about 18%. In the above described process, the new concept for the densification of powders before the partial melt process is used, namely a proper alternation of drawing and groove-rolling steps that drastically reduces the porosity and leads to a powder density inside the filaments much higher than that obtained by the standard process with just a drawing deformation. After the deformation process, different temperature treatments are used for each sample. Sample W_O2 must be considered the reference sample of the series, having undergone the standard thermal treatment in 1 bar oxygen atmosphere, consisting of a melting stage at 888 °C, a crystallization stage between 888 °C and 836°C with steps of different cooling rates (see Fig. 1), followed by a two-days temperature annealing plateau at 836 °C and a final slow cooling at 80 °C/h rate. Sample W_air has undergone the same heat treatment stages, but in ambient atmosphere, i.e. 21% oxygen partial pressure. Samples W_Q and W_Qbp are treated like the reference sample in oxygen atmosphere up to a certain stage, then the temperature is quenched to room temperature in air, in particular for sample W_Q there is no final slow cooling at 80 °C/h, while for sample W_Qbp both the temperature annealing plateau at 836 °C and the final slow cooling are omitted. We point out that this quenching procedure is not as fast as the quenching into room-temperature brine [20], but it is equally effective in stopping the sample oxygenation, as evidenced by Tc values presented in the following. Sample W_PA is the most oxygenated of the series, having undergone the standard thermal treatment in oxygen atmosphere like the reference sample, plus an additional post-annealing step at 400 °C for 30 h in oxygen atmosphere, followed by a slow cooling at 80 °C/h. Finally, sample W_air_Qbp is treated like sample W_Qbp, but in ambient atmosphere. In Table I, we list some measured parameters of the samples, namely magnetic transition temperatures $T_c$, transport critical current density at T=4 K and B=0, and brief descriptions of the fabrication protocols, including temperature treatment temperatures and durations, cooling and heating rates, atmosphere conditions. Simple sketches of the fabrication protocols are also shown in Figure 1.

| Sample name | $T_c$ (K) | $J_c$(T=4 K, B=0) (A/mm$^2$) | Thermal treatment |
|---|---|---|---|
| W_O2 | 81 | 3650 | Complete thermal treatment in $O_2$ atmosphere, including final slow cooling (80 °C/h) |
| W_air | 83 | 540 | Complete thermal treatment in air (21% $O_2$), including final slow cooling (80 °C/h) |
| W_Q | 90 | 3120 | Thermal treatment in $O_2$ atmosphere, without final slow cooling (80 °C/h) |
| W_Qbp | 90 | 2510 | Thermal treatment in $O_2$ atmosphere, without annealing plateau (836 °C for 48 h) and final slow cooling (80 °C/h) |
| W_PA | 78 | 4250 | Complete thermal treatment in $O_2$ atmosphere, including final slow cooling (80 °C/h) and post annealing (400 °C for 30 h) |
| W_Q_air_bp | 94 | 61 | Thermal treatment in in air (21% $O_2$), without annealing plateau (836 °C for 48 h) and final slow cooling (80 °C/h) |

**Table I:** list of samples with their magnetic onset $T_c$, zero field transport critical current density $J_c$ measured at T=4 K and details of the fabrication protocol.

**Results and discussion**

## a. Microstructural characterization

The relationship between microstructure and transport behavior in Bi-2212 wires is a major issue, which is addressed by an extensive amount of literature [7,8,9]. In this work, we propose to take into consideration a further aspect, i.e. the oxygenation of grain boundaries and its influence on supercurrent transport, therefore an in-depth analysis of the microstructure is beyond the scope of the present manuscript. Yet, an overall idea of the sample microstructure is obtained from SEM images, showing little differences from sample to sample in terms of grain size and shape. Also grain stoichiometry obtained from EDX analysis indicates the expected stoichiometry for our samples. Any information on the nature of GBs, involving the nanoscale, cannot be obtained by the SEM resolution. In Figure 2, SEM micrographs of W_O2 and W_air samples are shown. As seen in Table I, these samples have very different critical current density values and represent sharply different situations in our forthcoming $J_c$ analysis. Yet, they share quite similar morphology, made of platelet-like grains of tens of micron size. In a multi-element compound such as Bi-2212, the composition of GBs and grains may be mutually dependent, as phase segregation at the GBs may significantly alter the stoichiometry of grains. Oxygen stoichiometry of GBs may further complicate this interplay, as the condition of charge neutrality at GBs must be fulfilled. The strength of this interplay between grain and GB stoichiometries depends on the grain size, or more exactly, on the ratio of GB area to grain volume (i.e. specific area of GBs) [21,22] and must be taken into account for small enough grains, for which the specific area of GBs exceeds a threshold value (from measurements of solubility limit it turns out that this effect is negligible for grain size larger than ~1 micron and starts to be relevant for grain size smaller than ~100nm [21,22]). From our SEM micrographs, we find that the grain size is tens of microns, hence we can safely assume that the different $T_c$ values and $J_c(H)$ behaviors presented here must be related to mechanisms other than the sample-to-sample difference in elemental deficiency inside the grains, consequent to different secondary phase segregation at the GBs.

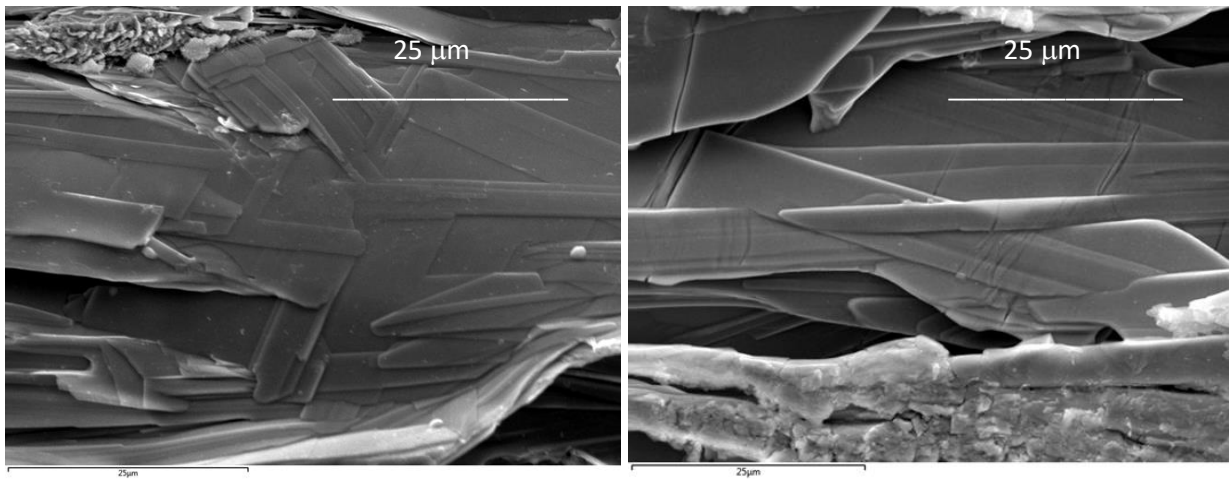

**Figure 2:** SEM images of W_O2 (left) and W_air (right) samples. The scale bars indicate a 25 microns length.

## b. Magnetic measurements

In Figure 2 we present zero field cooled magnetic moment curves, normalized to the 5 K value, measured as a function of temperature in an applied field of 10 Oe. Although the transition width are pretty broad, the critical temperature $T_c$ values are precisely extracted from the onset of the diamagnetic signal and range between 78 K and 94 K, as reported in Table I, the latter being the value of the W_Q_air_bp sample, the closest to optimal doping (we can assume $T_c^{opt} \approx 96$ K [23], even if $T_c^{opt}$ of Bi-2212 is composition-dependent and for our case $Bi_{2.16}Sr_{1.93}Ca_{0.89}Cu_{2.02}O_x$ it maybe 1-2 K lower). As this magnetic moment is generated by intra-grain currents, we can directly relate the $T_c$ values to the intra-grain oxygen doping in the different samples. From a comparison of heat treatment protocols, it is clear that the W_Q_air_bp sample must be the one with the lowest oxygen content, hence we gather that all the other samples are in the overdoped regime, from slightly overdoping (quenched samples) to heavy overdoping (post-annealed sample).

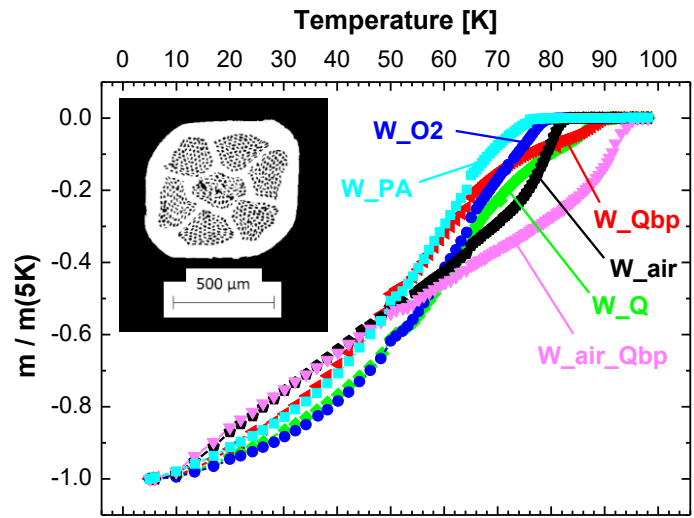

**Figure 3: Zero field cooled magnetic moment**. Zero field cooled magnetic moment curves versus temperature for the samples presented in this work, measured in a 10 Oe field. Inset: wire cross-section, which is the same for all the samples.

In the effort of putting this issue in more quantitative terms, in the upper panel of Figure 4 we plot the $T_c/T_c^{opt}$ values of the samples, overlapped to the characteristic dome of Bi-2212 [24] which also allows to map each $T_c$ to the number of holes per Cu plane and thus to oxygen doping. From this result it comes out that the crystallization step down to 836 °C is effective in doping grains with oxygen near to the optimal content, even in an atmosphere only 21% oxygen partial pressure (W_air_Qbp sample). If this crystallization step is carried out in oxygen atmosphere, a moderately overdoped regime of grains is achieved (W_Q and W_Qbp). The temperature plateau at 836 °C has no effect on the intra-grain oxygenation (compare $T_c$'s of W_Q and W_Qbp). Successive treatments, namely slow cooling and post annealing, further enhance the degree of grain overdoping (W_air , W_O2 and W_PA), more efficiently if carried out in oxygen atmosphere (W_O2 and W_PA).

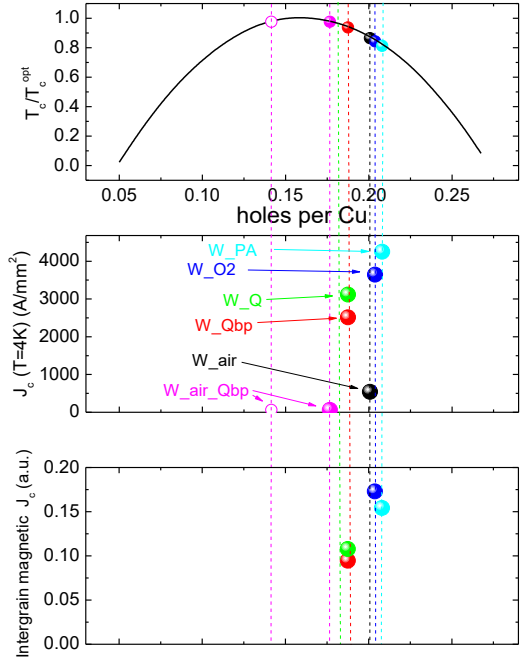

**Figure 4: Critical temperature and inter-grain critical current density plotted as a function of intra-grain oxygen doping.** Upper panel: $T_c/T_c^{opt}$ values of the samples plotted on the Bi-2212 dome to map the number of holes per Copper plane associated to intra-grain oxygen doping [24]. Sample W_air_Qpb is placed at both sides of the dome peak because it is the closest to optimal doping but we cannot tell whether it lies at the overdoped or underdoped side. Middle panel: transport critical current $J_c$ values measured at 4 K for the same samples, plotted along the same horizontal axis, which clearly shows that $J_c$ values are not correlated to intra-grain behavior. Lower panel: estimation of inter-grain $J_c$ extracted form remnant magnetization data, plotted along the same horizontal axis. In particular the values of $m_R$ in emu/g at the low-$H_{max}$ peak are reported (see text and Figure 5), which should represent relative inter-grain $J_c$ values if the same macroscopic scale of the current is assumed for all the samples. These magnetization inter-grain $J_c$ are consistent with transport $J_c$ data in the middle panel.

For a preliminary comparison, in the middle panel of Figure 3 we show transport critical current $J_c$ values measured at 4 K and self-field for the same samples, also reported in Table I, plotted along the same horizontal axis as intra-grain doping. We can anticipate that these transport $J_c$ values are substantially consistent with data extracted from remnant magnetization (lower panel of Figure 3), yet clearly completely uncorrelated to intra-grain behavior (upper panel of Figure 3), suggesting that they are rather determined by inter-grain current and thus grain boundary properties. The largest transport $J_c$=4250 A/mm$^2$ is measured in the post annealed W_PA sample, but a similarly high value $J_c$=3650 A/mm$^2$ is measured in the reference W_O2 sample. The $J_c$ values of the quenched samples are still fairly high, namely 3120 A/mm$^2$ and 2510 A/mm$^2$ for W_Q and W_Qbp, respectively. As for the samples treated in ambient atmosphere W_air and W_air_Qbp, much smaller values 540 A/mm$^2$ and 61 A/mm$^2$ are obtained. These results will be discussed later on, while also inspecting the temperature dependence of the $J_c$ curves.

In Figure 5 we present remnant magnetization data of the series of samples, plotted as the derivative of the remnant magnetization with respect to the logarithm of the field as a function of the maximum field reached in the magnetization cycle $H_{max}$. Although numerical differentiation introduces some scattering, it can be clearly seen that for some samples, namely those with large transport $J_c$, the curves exhibit a double peak shape, while for other samples, the ones with very low transport $J_c$, only the peak at the largest $H_{max}$ field survives. As mentioned in the experimental section above, the peak at the lowest $H_{max}$ field is associated to magnetic flux first entering grain boundaries weak links and the value of $m_R$ at this peak is proportional to the inter-grain magnetic critical current density. If we assume that the macroscopic scale of inter-grain current is equal for all the samples, we can compare the values of the peak $m_R$, as done in the lower panel of Figure 3. It can be seen that, in agreement with transport data, W_O2 and W_PA samples have the largest $J_c$, while W_Q and W_Qbp samples have lower but still considerable $J_c$. No magnetic $J_c$ estimation is possible for W_air and W_air_Qpb samples.

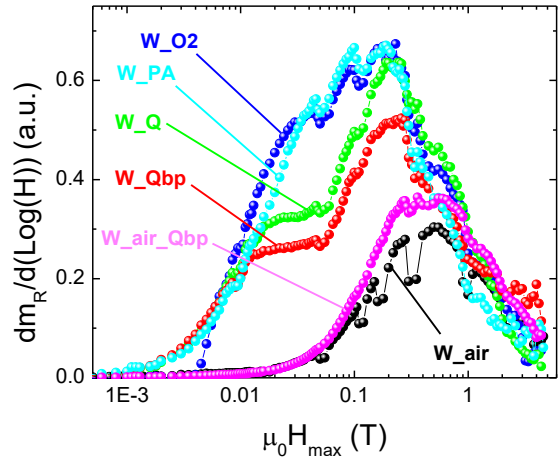

**Figure 5: Remnant magnetization**. Remnant magnetization plots of the series of samples, plotted as the derivative of the remnant magnetization $m_R$ with respect to the base 10 logarithm of the field as a function of the maximum field reached in the magnetization cycle $H_{max}$.

**c. Transport critical current measurements**

In Figure 6, the transport $J_c(T)$ curves measured in the reference W_O2 sample at different fields are shown. The temperature and field dependence of $J_c$ of this sample is representative of all the series, when normalized to the absolute value at the lowest temperature and zero field. In the upper panel it can be seen that in zero field $J_c$ decreases quite rapidly from 3650 A/mm$^2$ to 1910 A/mm$^2$ with increasing temperature from 4 K to 24 K, even if this temperature regime is still much lower than $T_c$. Even larger suppression of $J_c$ with temperature is measured in applied field. A scaling based on collective pinning theory [14,25,26] is usually applied to describe Bi-2212 $J_c(T,H)$ curves:

$$J_c(T,B) = J_{sc}(T) \exp\left(-\frac{B}{B_{sc}(T)}\right) \quad (1)$$

where $B_{sc}$ and $J_{sc}$ are scaling parameters which obey the following temperature dependence, with $T_{sc}$ as additional parameter:

$$B_{sc}(T) = B_{sc}(0) \exp\left(-\frac{T}{T_{sc}}\right) \quad (2)$$

$$J_{sc}(T) \approx J_{sc}(0)\left(1-\frac{T}{T_{sc}}\right)^{\alpha} \qquad (3)$$

The scaling is shown in the lower panel of Figure 6 as continuous lines, and its validity holds only for fields ≥2 T, suggesting that at lower fields the flux vortices do not behave in a collective mode, as the vortex-vortex spacing is such that the interaction between each vortex line and the pinning centers is stronger than the vortex-vortex interaction. The values of the fitting parameters are consistent with the literature [14,26], for example for W_O2 sample we find $B_{sc} \approx 60$ T and $T_{sc} \approx 8$ K.

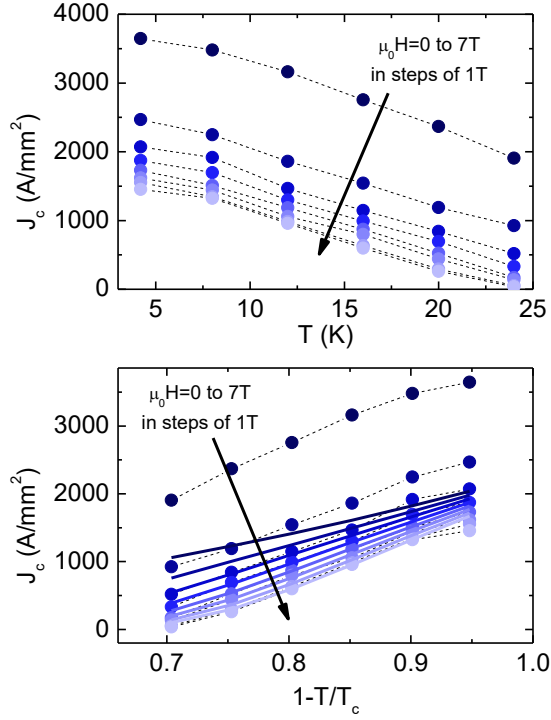

**Figure 6: Temperature dependence of Jc at different fields.** Temperature dependence of $J_c$ at different fields for the W_O2 reference sample, plotted versus T (upper panel) or $1-T/T_c$ (lower panel), and fitting curves based on collective pinning theory (see text), valid at fields ≥ 2 T, plotted as thick continuous lines. In this figure, the color scale identifying magnetic field values is the same for symbols representing experimental data and for thick continuous lines representing fitting curves.

We now focus of the temperature dependence of $J_c$ at zero field in the different samples, which is rich in information regarding weak link behavior and proximity effect at the grain boundaries. In Figure 7 we plot $J_c(T)$ of the different samples.

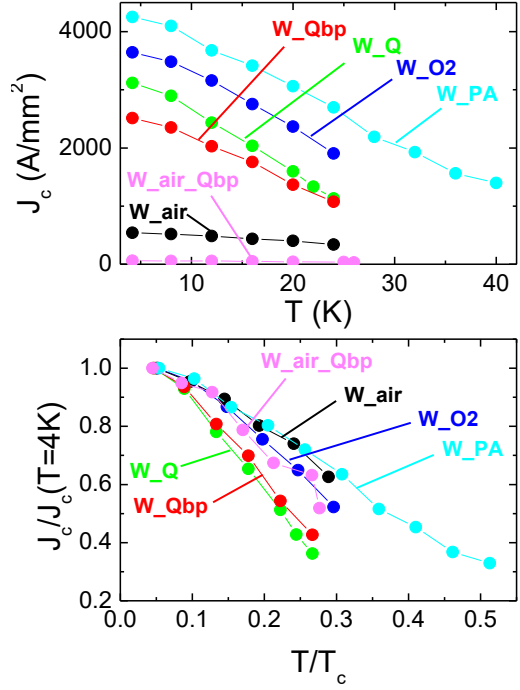

**Figure 7: Temperature dependence of Jc in different samples.** Temperature dependence of $J_c$ (upper panel) or $J_c$ normalized at 4 K (lower panel) in zero field for the different samples, plotted in terms of absolute (upper panel) or reduced (lower panel) values. Lines are guides to the eye.

In order to analyze these curves, the $J_c$ temperature dependence can be compared with a law of the type $J_c \sim J_{c0}(1-T/T_c)^{\alpha}$, as in eq. (3), with $J_{c0}$ and $\alpha$ as fitting parameters. The value of the exponent $\alpha$ is predicted to be $\alpha=2$ for an array of superconductor-normal metal-superconductor (SNS) Josephson junctions [27], and $\alpha=1$ for an array of superconductor-insulator-superconductor (SIS) Josephson junctions [27][28]. However, such models strictly apply as long as T is not much smaller than $T_c$ as they are developed within the Ginzburg-Landau expansion [27], and moreover they do not consider the proximity effect potentially occurring at S/N boundaries. Indeed, if we try to fit data with these models, even disregarding data at the lowest temperatures, we obtain exponent values ranging from 1.6 to 3.9 among the different samples, clearly out of the range predicted by the models. The proximity effect, whose theory was first formulated by de Gennes [27], reduces the gap in a superconductor close to the interface with a normal metal within a layer of thickness $\xi_S$ from the interface and induces a finite gap in the normal metal within a layer of thickness $\xi_N$ from the interface, being $\xi_S$ and $\xi_N$ the coherence lengths in the superconductor and in the normal metal, respectively. $\xi_N$ is typically 2-3 nm for non-superconducting oxides akin to high-$T_c$ superconducting cuprates (HTSCs) (see Table II in [29]). It is therefore more convenient to compare our $J_c(T)$ curves with existing models that take into account the proximity effect, such as the models for individual tunnel junctions (SIS) and weak links (SNS), developed in ref. [30], valid throughout the entire temperature range below $T_c$. This work calculates the $J_c(T)$ dependence across an individual weak link on the basis of the Usadel equations for dirty superconductors [31], whose validity extends in the whole temperature range, rather than on the Ginzburg–Landau relations used by de Gennes. In this comparison, we disregard the fact that our polycrystalline samples are actually arrays of junctions rather than an individual junction.

In the formulation by Golubov and Kupriyanov [30], the spatial dependence of the Cooper pair order parameter across a SN weak link is calculated by solving the analytically-continued Usadel equations for the modified Usadel functions $\Phi_{N,S}$ (see eq. (3) in [30]), with suitable boundary conditions at the

SN interface and at asymptotic distance from it (see eqs. (4-8) in [30]). In these equations and boundary conditions, two key parameters describe the junction, namely $\gamma_B$ and $\gamma_M$. The former, $\gamma_B$, is an interface parameter describing the boundary transparency (decreasing $\gamma_B$ means increasing transparency and thus increasing $J_c$ of the junction, especially in the range $0<\gamma_B<1$). It is expressed as:

$$\gamma_B = \frac{R_B}{\rho_N \xi_N} \frac{d_N}{\xi_N} \qquad (4)$$

where $R_B$ is the product of the resistance and area of the S/N boundary, $\rho_N$ is the resistivity of the N barrier, $d_N$ is the thickness of the N barrier. On the other hand, the parameter $\gamma_M$ is a measure of the suppression of the order parameter due to the proximity effect ($\gamma_M=0$ means no proximity effect) and it is expressed as:

$$\gamma_M = \frac{\rho_S \xi_S}{\rho_N \xi_N} \frac{d_N}{\xi_N} \qquad (5)$$

In ref. [30] it is shown that calculation of the $\Phi_N$ and $\Phi_S$ functions as well as of the state densities in the N layer makes it possible to calculate the tunnel current in SNS junctions. Unfortunately, analytical solutions are found only for limiting cases of temperature and parameters space. For example, the limiting case of a SIS junction with zero proximity effect and ideal (maximum) transparency has $\gamma_B=0$ and $\gamma_M=0$, which corresponds to the Ambegaokar-Baratoff (AB) formula [28]. Numerical techniques are used in ref. [30] to solve the equations for arbitrary temperatures and values of $\gamma_B=0$ and $\gamma_M=0$. Before comparing the predictions of the Golubov-Kupriyanov model with our samples, we make some consideration about its applicability and boundary conditions. In HTSCs rigid boundary conditions are appropriate [29], which means conditions of extreme mismatch in transport properties between the S and N materials, so that the suppression of the order parameter at the S/N boundary from the S side is almost negligible. As for the relevance of proximity effect at HTSCs grain boundaries, it may be questioned, indeed in most artificial junctions of different type a linear temperature dependence of $J_c$ is experimentally found, which is considered as an evidence of absence of proximity effect [29]. However, it was remarked that such artificial junctions may be affected by fabrication problems such as pinholes; by converse, in junctions whose barrier is made of a chemically substituted form of the same compound as that of superconducting HTSC electrodes, $J_c(T)$ was found to have an exponential temperature dependence, which means that proximity effect is relevant [29]. The case of grain boundaries in bulk HTSCs seems to be analogous to this latter case, hence the Golubov-Kupriyanov equations including proximity effect ($\gamma_B \neq 0$ and $\gamma_M \neq 0$) should be applicable.

After these premises, given that the general results obtained via the Usadel equations are not simple analytic expressions, we compare our experimental $J_c(T)$ curves, normalized to their low temperature limit $J_c(T=4\ K)$, with the ones calculated numerically for different values of the parameters, as found in references [30,29,32]. In the following discussion we focus on the central role of the parameter $\xi_N$, upon which the tunneling current depends exponentially, as compared to other parameters such as $R_B$ and $\rho_N$ in eq. (4) and (5), that appear as pre-factors in the expression of the tunneling current [29].

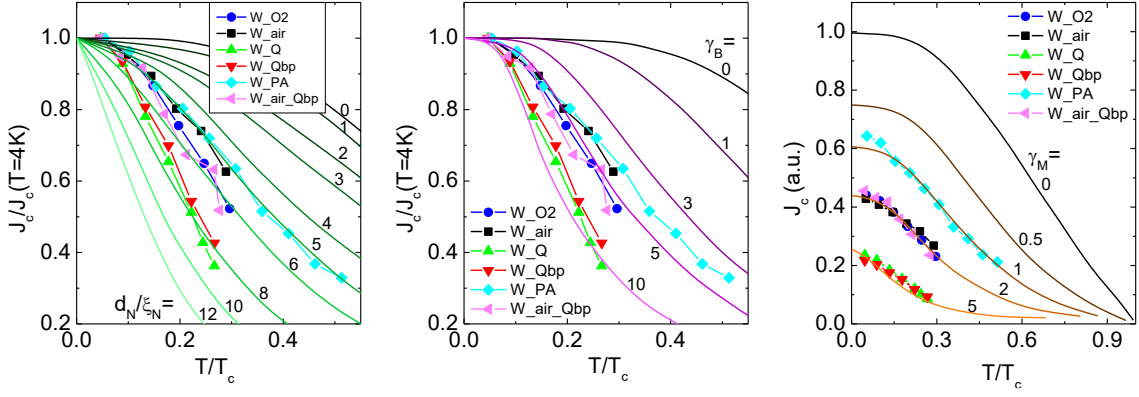

**Figure 8: Temperature dependence of normalized Jc, compared to predictions of models for individual tunnel junctions and weak links.** a) normalized critical current plotted versus the reduced temperature $T/T_c$ for different values of the ratio $d_N/\xi_N$ from zero to 12 (top to bottom), as calculated in [29] using Likharev's theory of SNS junctions in the dirty limit under rigid boundary conditions, which are the regimes relevant to HTSCs. b) and c) normalized critical current plotted versus the reduced temperature $T/T_c$ for different values of the parameters $\gamma_B$ and $\gamma_M$, as calculated in [30] for a SNINS junction, in particular in b) $\gamma_M=0$ and $\gamma_B=0, 1, 2, 3, 5, 10$, while in c) $\gamma_B=1$ and $\gamma_M=0, 0.5, 1, 2, 5$. In c), the calculated curves are normalized to $J_c(\gamma_M=0, T=0)$ and the experimental curves are scaled by an arbitrary factor, which on the other hand does not prevent to identify univocally which of the calculated curves better describes the temperature dependence.

In Figure 8a), the computed normalized critical current is plotted versus the reduced temperature $T/T_c$ for different values of the ratio $d_N/\xi_N$ from zero to 12, as calculated on the basis of Likharev's work [32,33]. It is seen that $J_c$ is severely suppressed with increasing temperature as the ratio $d_N/\xi_N$ increases, indeed $J_c$ depends exponentially on $d_N/\xi_N$ [32]. If we plot on the same graph our experimental normalized $J_c(T)$ curves, we find that curve of each sample nearly overlaps with one of the calculated curves, which identifies the parameter values that better describe the respective sample. This comparison of experimental data with calculated curves indicates clearly that the ratio $d_N/\xi_N$ increases in the W_Q and W_Qbp samples as compared to the reference W_O2 sample from 5-6 to 8. Assuming that the barrier thickness is roughly the same in the different samples, we can hypothesize that $\xi_N$ changes in the W_Q and W_Qbp samples as compared to the W_O2 sample. $\xi_N$ is expected to decrease with increasing resistivity and decreasing carrier density in the barrier as $\sim \rho_N^{-1/3}$ and $\sim n^{1/3}$ [i]. Thus, if the barrier layer changes from metallic towards insulating, the accompanying resistivity increase results in a smaller coherence length $\xi_N$ [29]. Hence, this indicates that the quenching process has a sizeable effect not only on grains, driving them from the highly overdoped toward less overdoped region of the phase diagram (see upper panel of Figure 4), but also on grain boundaries, making them more resistive and thus less transparent (likely as a consequence of oxygen doping [13], but possibly also due to stoichiometry or microstructural changes). We point out that also the comparison of $J_c(T=4\text{ K})$ absolute values of quenched and reference samples confirms this scenario, as W_Q and W_Qbp samples have lower, though still fairly large, $J_c(T=4\text{ K})$ values as compared to the reference W_O2 sample. Moreover, the fact that W_Q has a larger $J_c(T=4\text{ K})$ than W_Qbp

---

[i] By recalling that in the Drude model Fermi velocity $v_F$, mean free path $\ell$ and diffusivity can be estimated with the expressions $v_F = \frac{\hbar(3\pi^2 n)^{1/3}}{m}$, $\ell = \frac{m v_F}{e^2 n \rho_N}$ and $D \propto v_F \ell$, respectively, where n is the carrier density, m is the effective mass, $e$ is the electron charge, $\hbar$ is the Planck constant, $\rho_N$ is the resistivity, it turns out that in the dirty limit ($\ell \ll \xi_N$), $\xi_N = \sqrt{\frac{\hbar}{2\pi k_B T} D} \propto \rho_N^{-1/2} n^{-1/6} \propto \rho_N^{-1/3} \propto n^{1/3}$, where $k_B$ is the Boltzmann constant

suggests that the plateau at 836 °C is either beneficial for the oxygenation of grain boundaries or favors grain growth, thus decreasing the effective number of grain boundaries crossed by the transport supercurrent (although the latter effect is plausible, we do not have clear evidence of this by our SEM analyses, which should be carried out in samples quenched just before and after the plateau).

A different trend is observed for the W_PA sample, where the ratio $d_N/\xi_N$ decreases from 5-6 to 4 as compared to the reference sample W_O2, suggesting that $\xi_N$ increases, as a consequence of the decreased resistivity and increased carrier density in the barrier. This demonstrates that the post-annealing treatment in $O_2$ not only drives the grains toward even more strongly overdoped regime, but it also dopes the grain boundaries, making them more metallic and thus more transparent. Also in this case the analysis of the $J_c(T)$ dependence is consistent with the comparison of $J_c(T=4 K)$ absolute values of post annealed and reference samples, being the $J_c(T=4 K)$ of the former larger. However, it must be remarked that this $J_c$ gain is obtained with an onerous time consuming process, namely a 30 h annealing, and large scale fabrication needs to balance the pros and cons in these terms.

Finally, in Figure 8a), the W_air sample behaves similarly to the W_O2 sample in terms of grain boundary transparency. This indicates that, during heat treatments, the conducting grain boundaries are doped with oxygen from the atmosphere even with only 21% oxygen partial pressure. However, in this case the comparison of $J_c(T=4 K)$ absolute values of samples prepared in ambient or oxygen atmosphere shows that the samples prepared in air have significantly suppressed $J_c$. This apparent inconsistency may be due to the fact that if thermal treatment is carried out in air, several spurious phases form at the grain-boundaries, which totally obstruct the supercurrent flow (at odds with conducting grain boundaries that behave as weak links of larger or smaller transparency to supercurrents, depending on their oxygen doping). Such spurious phases thus shrink the effective cross section of conducting grain boundaries to the supercurrent. This hypothesis is supported by phase studies [34] [35] [36], where $(Sr,Ca)CuO_2$ and $Bi_2(Sr,Ca)_4CuO_x$ were identified as secondary phases in air treated samples. Indeed in [34] it was shown that $(Sr,Ca)CuO_2$ large crystals are the main non-superconducting phase present in air treated samples as the 2212 phase forms from the melt. These large crystals do not fully react to form 2212 during cooling and annealing, thus determining a poor microstructure in fully processed tape, that obstructs $J_c$. Yet, more recently, the influence of the oxygen partial pressure on the phase changes that occur during Bi-2212 melt processing of a state-of-the-art Bi-2212 multifilamentary wire was monitored *in situ* by high energy synchrotron X-ray diffraction [37] and it was found that only tiny amounts of secondary $Bi_2(Sr,Ca)_4CuO_x$ phase are present in air treated samples. The latter result, together with the similar $J_c(T)$ dependence of W_air and W_O2 samples, made us trustful about the validity of comparing grain boundary properties of samples heat treated in air with those of samples treated in oxygen.

In Figure 8b) and c), the computed normalized current is plotted versus the reduced temperature $T/T_c$ for different values of the parameters $\gamma_B$ and $\gamma_M$, as calculated in ref. [30], in particular in Figure 8b) $\gamma_M=0$ and $\gamma_B$ varies from zero to 10, while in Figure 8c) $\gamma_B=1$ and $\gamma_M$ varies from 0 to 5. In some HTSC junctions, it was found that $\gamma_M<<\gamma_B$ and $\gamma_M>1$ [30], thus likely the plots in Figure 8b) better apply to our samples. In any case, regardless the particular choice of parameters values, the comparison of experimental curves with calculated ones unambiguously indicates that in the W_Q and W_Qbp samples either the parameter $\gamma_B$ or the parameter $\gamma_M$, or else both of them, increase as compared to the reference W_O2 sample. Considering the expression (4) and (5) for $\gamma_B$ and $\gamma_M$, this trend can be again explained with a decrease of $\xi_N$ in the W_Q sample as compared to the W_O2 sample, related to more insulating character of grain boundaries. On the other hand, in the W_PA sample either the parameter $\gamma_B$ or the parameter $\gamma_M$, or else both of them, decrease, as a consequence of the increase of $\xi_N$ in the more metallic grain boundaries.

We remark that in the above analysis and discussion we disregard the fact that $\xi_s$ the coherence length in the superconducting grains may vary from sample to sample, similarly to $\xi_N$, the coherence length in the normal phase at the grain boundaries. Moreover, on the basis of SEM analyses, we rule out that the grain size may also vary significantly from sample to sample, thus changing the actual number of grain boundaries that the supercurrent crosses. However, the overall scenario seems to catch the main aspects and allows to clarify the role of each heat treatment stage on the doping of the grains and of the grain boundaries.

**Conclusions**

We present a full characterization of Bi-2212 wires prepared with different protocols, purposely chosen to investigate the effect of oxygen doping on the grain boundary transparency to current and on the superconducting properties in general. We examine $T_c$ values and $J_c$ behavior as a function of temperature from 4 K to 24 K and magnetic field up to 7T. As for intra-grain oxygenation, inferred from $T_c$ values, the crystallization step down to 836 °C is effective in doping grains with oxygen near to the optimal content in ambient atmosphere and to the moderately overdoped regime in oxygen atmosphere. The temperature plateau at 836 °C has no effect on the intra-grain oxygenation, while successive slow cooling and post annealing treatments further enhance the degree of overdoping, especially if carried out in oxygen atmosphere. As for grain boundary oxygenation, inferred from the analysis of the $J_c(T)$ dependence, that allows to extract information about the weak link behavior and proximity effect at the grain boundaries, we gather that it takes place partially in the crystallization stage, but it becomes complete in the plateau at 836 °C and in slow cooling stages. A further grain boundary oxygenation occurs in the prolonged post annealing step.

In general, the analysis of the temperature dependence of $J_c$ is an investigation method that could be applicable to all superconductors of applicative interest, whose weak links are an obstacle to supercurrent (e.g. cuprates, iron pnictides and chalcogenides).

**Ancknowledgemnts**

The authors would like to thank Cristina Bernini for the SEM images.